\documentclass[aps,pra,10pt,twocolumn,nofootinbib,superscriptaddress]{revtex4-1}
\usepackage[latin1]{inputenc}
\usepackage[english]{babel}
\usepackage{graphicx}
\usepackage{color}
\usepackage{amsmath}
\usepackage{amssymb}
\usepackage{multirow}
\usepackage{color}

\graphicspath{{./figs/}}

\begin{document}
	
 \title{Modeling anisotropic magnetized strange quark stars}
 
 \author{S. L\'opez P\'erez\footnote{samantha@icimaf.cu}}
 \email{samantha@icimaf.cu}
 \affiliation{Facultad de F{\'i}sica, Universidad de la Habana,\\ San L{\'a}zaro y L, Vedado, La Habana 10400, Cuba}
 
 \author{D. Manreza Paret}
 \affiliation{Facultad de F{\'i}sica, Universidad de la Habana,\\ San L{\'a}zaro y L, Vedado, La Habana 10400, Cuba}

 \author{G. Quintero Angulo}
 \affiliation{Facultad de F{\'i}sica, Universidad de la Habana,\\ San L{\'a}zaro y L, Vedado, La Habana 10400, Cuba}
  
 \author{A. P\'erez Mart\'{\i}nez}
 \affiliation{Instituto de Cibern\'{e}tica, Matem\'{a}tica y F\'{\i}sica, \\
 	Calle E esq a 15, Vedado 10400, La Habana Cuba}
 
 \author{D. Alvear Terrero}

 \affiliation{Institute of Theoretical Physics, University of Wroclaw, \\
 	pl. Maxa Borna 9, 50-204, Wroclaw, Poland}

\begin{abstract}
 Compact objects have an intrinsic anisotropy due to the presence of strong magnetic fields that cause considerable modifications  on the equations of state (EoS). In this work, we study the impact of this anisotropy in the size and shape of magnetized strange quark stars using an axially symmetric metric in spherical coordinates, the gamma-metric. The results are compared with those obtained with the standard  Tolman-Oppenheimer-Volkoff (TOV) equations for the parallel and perpendicular pressures independently. Differences in the results are discussed.
 
\end{abstract}

\maketitle
 

 \section{Introduction}\label{sec1}
 
 It is hypothesized that  individual neutrons in a neutron star  under a regimen of strong gravity, break down into their constituent quarks
 (up and down), forming what is known as quark matter.  A specific kind of quark matter is known as strange quark matter (SQM) and it is formed when quarks up and down transform into strange quarks. SQM is speculated to be the stable ground state of strong-interaction matter (Bodmer-Witten's conjecture)~\citep{Bodmer1971, Witten1984}. This is only possible for high densities, so the most likely place to find strange quark matter in nature would be inside neutron stars cores. Quark stars made of strange quark matter are called strange stars (SSs) and were first proposed by ~\citep{Itoh} when he suggested the idea of a star composed by 3-flavour quark matter. Nowadays, astronomers are still searching for evidence of strange stars. Observations released by the Chandra X-ray Observatory detected two possible candidates: RX J1856.5-3754 and 3C58 \citep{Drake:2002bj}.
 
 On the other hand, magnetic fields are present in almost all stars during their evolution, becoming huge in their final stage. In the case of neutron stars, observed surface
 magnetic fields range from $10^{9}$ to $10^{15}$ G \citep{1992ApJ...392L...9D,1998Natur.393..235K}, while their inner magnetic fields might be as high as $5\times10^{18}$ G ~\citep{Daryel22015}. Although
 the inner magnetic fields cannot be observed directly, their
 bounds can be estimated with theoretical models based on macroscopic and microscopic considerations \citep{Lattimerprognosis}.
 
 A magnetic field acting on a fermion gas breaks the spherical symmetry and produces an anisotropy in the quantum-statistical average of the energy-momentum tensor. The effect of this anisotropy is the splitting of the pressure into two components, one along the magnetic field \textemdash the parallel pressure $P_{\parallel}$\textemdash  and another in the transverse direction \textemdash the perpendicular pressure $P_{\perp}$\textemdash, so that $T^{\mu}_{\nu}=\text{diag}(-E, P_{\perp},P_{\perp},P_{\parallel})$. Consequently, a gas of fermions under the action of a constant and uniform magnetic field has an anisotropic\textemdash axially symmetric\textemdash equation of state~\cite{Chaichian}. For this reason, when modeling the structure of magnetized compact objects, one should consider axial symmetry instead of the spherical symmetry used when solving the Tolman-Oppenheimer-Volkoff (TOV) equations.
 
 In this work, we compare the size and shape of magnetized strange quark stars using two different sets of structure equations, to study the magnetic field effects in their masses and radii. In Sections \ref{sec2} and \ref{sec3} we revisited the studies for magnetized SQM performed in \citep{Felipe:2008cm}. Section \ref{sec2} presents magnetized SSs EoS and discuss the magnetic field effects on the energy density and pressure while section \ref{sec3} shows TOV solutions. The anisotropic structure equations \citep{prd} are presented in Section \ref{sec4} with their corresponding numerical results for magnetized strange stars. Concluding remarks are given in Section \ref{sec6}.
 
 
 \section{EOS for magnetized strange quark stars}\label{sec2}
 
 A pure SS is a compact object (CO) exclusively composed by strange quark matter and electrons. To describe quarks in the interior of the star we use the phenomenological MIT Bag model \citep{Chodos}. In this model, quarks are assumed as quasi-free particles confined into a \textquotedblleft bag\textquotedblright\, that reproduces the asymptotic freedom and confinement through the B$_{\text{bag}}$ parameter. In addition, the star is under the action of a uniform and constant magnetic field oriented in the $z$ direction ${\bf B} = (0,0,B)$.
 
 The pressure and the energy density of the magnetized gas of quarks and electrons, are obtained from the thermodynamical potential\citep{Felipe:2007vb}
 
 \begin{multline}\label{Thermo-Potential}
 \Omega_f(B,\mu_f,T)  =   -\frac{e_fd_f B}{\beta}  \!\!\!  \int_{-\infty}^\infty \! \frac{dp_3}{4\pi^2}  \! \sum_{l=0}^{\infty}g_l \\
 \times\sum_{p_4}\ln\bigg[(p_4+i\mu_f)^2+ \varepsilon^2_{lf} \bigg],
 \end{multline}
 where $l$ stands for the Landau levels, $d_f$ is the flavour degeneracy factor\footnote{$d_e=1$ and $d_{u,d,s}=3$} and $g_l=2-\delta_{l0}$ include the spin degeneracy of the fermions for $l\neq0$. Moreover, $\beta$ is the inverse of the absolute temperature $T$, $\mu_f$, $m_f$ and $e_f$  are the chemical potential, mass and charge of particles respectively. The spectrum of the  charged fermions in the magnetic field is $\varepsilon_{lf}=\sqrt{p_3^2+2|e_fB|l+m_f^2}$\footnote{$f=e, u, d, s$ stands for electron and each quark flavour}.  
 
 Eq. \eqref{Thermo-Potential} can be divided in two contributions:
 \begin{equation}\label{Thermo-Potential-2}
 \Omega_f (B,\mu_f,T)=\Omega^\text{vac}_f(B,0,0)+\Omega^\text{st}_f(B,\mu_f,T),
 \end{equation}
 with
 \begin{eqnarray}
 \Omega^\text{vac}_f(B,0,0) &=&-\frac{e_fd_f B}{2\pi^2}\int_{0}^{\infty}\hspace{-2mm}dp_3\sum_{l=0}^{\infty}g_l\varepsilon_{lf} \label{vaccon},\\
 \Omega^\text{st}_f(B,\mu_f,T) &=& -\frac{e_fd_f B}{2\pi^2\beta}\int_{0}^{\infty}\hspace{-2mm}dp_3\sum_{l=0}^{\infty}g_l 
 \ln\hspace{-1mm}\big[1+e^{-\beta(\varepsilon_{lf}\pm\mu_f)}\big]
 \label{stcon}\hspace{-0.5mm}.
 \end{eqnarray}
 The vacuum contribution in Eq. \eqref{vaccon} does not depend on the chemical potential nor on the temperature and presents an ultraviolet divergence that must be renormalized \citep{Schwinger51}, resulting in
 \begin{equation}\label{vaccon2}
 \Omega^\text{vac}_f(B,0,0)= \frac{d_fm_f^4}{24\pi^2}\left(\frac{B}{B^c_f}\right)^2\text{ln}\frac{B}{B^c_f}.
 \end{equation}
 In Eq. \eqref{vaccon2} $B^c_f=m^2_f/e_f$ is the critical magnetic field (Schwinger field)\footnote{The magnetic field at which the cyclotron energy of the particles is comparable to their rest mass.}. For electrons $B^c_e\sim10^{13}$~G while for quarks up, down and strange we have $B^c_u\sim10^{15}$ G, $B^c_d\sim10^{16}$ G and $B^c_s\sim10^{19}$ G, respectively.
 
 COs have temperatures much smaller than the Fermi temperature of the gases that compose them. Hence, a good approximation is to compute the thermodynamical potential of these gases in the degenerate limit (T$\rightarrow0$) \citep{Shapiro, Camenzind}. In that case, the statistical term becomes
 \begin{equation}\label{stcon2}
 \Omega^\text{st}_f(B,\mu_f,0)=-\frac{e_fd_f B}{2\pi^2}\int_{0}^{\infty}dp_3\sum_{l=0}^{\infty}g_l\Theta(\mu_f-\varepsilon_{lf}),
 \end{equation}
 where $\Theta(\zeta)$ is the unit step function. From  Eq.  \eqref{stcon2}, we obtain
 \begin{multline}\label{stcon3}
 \Omega^\text{st}_f(B,\mu_f,0)=-\frac{e_fd_f B}{4\pi^2}\sum_{0}^{l_{max}}g_l\bigg[\mu_fp_f^l \\
 -(\varepsilon_f^0)^2\text{ln}\bigg(\frac{\mu_f+p_f^l}{\varepsilon_f^0}\bigg)\bigg],
 \end{multline}
 where $p_f^l=\sqrt{\mu_f^2-(\varepsilon_f^0)^2}$, $\varepsilon_f^0=\sqrt{m_f^2+2qBl}$ and $l_{\text{max}}=I[\frac{\mu_f^2-m_f^2}{2e_fB}]$. $I[z]$ denotes the integer part of $z$.
 
 The vacuum contribution in Eq. \eqref{vaccon2} can be neglected when is compared to the statistical one in Eq.  \eqref{stcon3}, given the high fermionic densities.
 Therefore, the thermodynamical potential of the degenerate fermion system can be approximated to $\Omega_f(B,\mu_f,0)=\Omega^\text{st}_f(B,\mu_f,0)$.
 
 Strange quark matter inside the star must be in stellar equilibrium. So, we impose $\beta$ equilibrium, charge neutrality and baryon number conservation  to the system in terms of the particle density $N_f=-\partial\Omega_f/\partial\mu_f$ and the chemical potentials. These conditions are
 
 \begin{subequations}\label{eqsqm}
 	\begin{eqnarray}
 	\mu_u+\mu_e-\mu_d=0 \,\,\,, \,\,\, \mu_d-\mu_s&=& 0,\\
 	2N_u-N_d-N_s-3N_e&=& 0, \\
 	N_u+N_d+N_s-3N_B&=& 0.
 	\end{eqnarray}
 \end{subequations}
 
 With these considerations, the magnetized SSs EoS become
 \begin{subequations}\label{EoSSS}
 	\begin{align}
 	E &= \sum_{f}\left[\Omega_f+ \mu_fN_f\right]+B_{\text{bag}} +\frac{B^2}{8\pi},\label{EoS1}\\
 	P_{\parallel} &=- \sum_{f}\Omega_f-B_{\text{bag}}-\frac{B^2}{8\pi}, \label{EoS2}\\
 	P_{\perp} &= -\sum_{f}\left[\Omega_f+B\mathcal{M}_f\right]-B_{\text{bag}}+\frac{B^2}{8\pi},\label{EoS3}
 	\end{align}
 \end{subequations}
 where $\mathcal{M}_f=-\partial\Omega_f/\partial B$ is the magnetization. The last term in Eqs. \eqref{EoSSS} is the Maxwell contribution to the pressures and the energy density $P^B_{\perp}=E^B=-P^B_{\parallel}=B^2/8\pi$. We consider B$_{bag}=75\,$MeV~fm$^{-3}$ which guarantees the stability of SQM  at  $B=0$, for a strange quark mass of $150$~MeV and baryon density of $2.18$~fm$^{-3}$ \citep{Felipe:2008cm}. A more realistic approach should also take into account the B$_{bag}$ dependence with the magnetic field.
 
\begin{figure}[!h]
 	\includegraphics[width=0.48\textwidth]{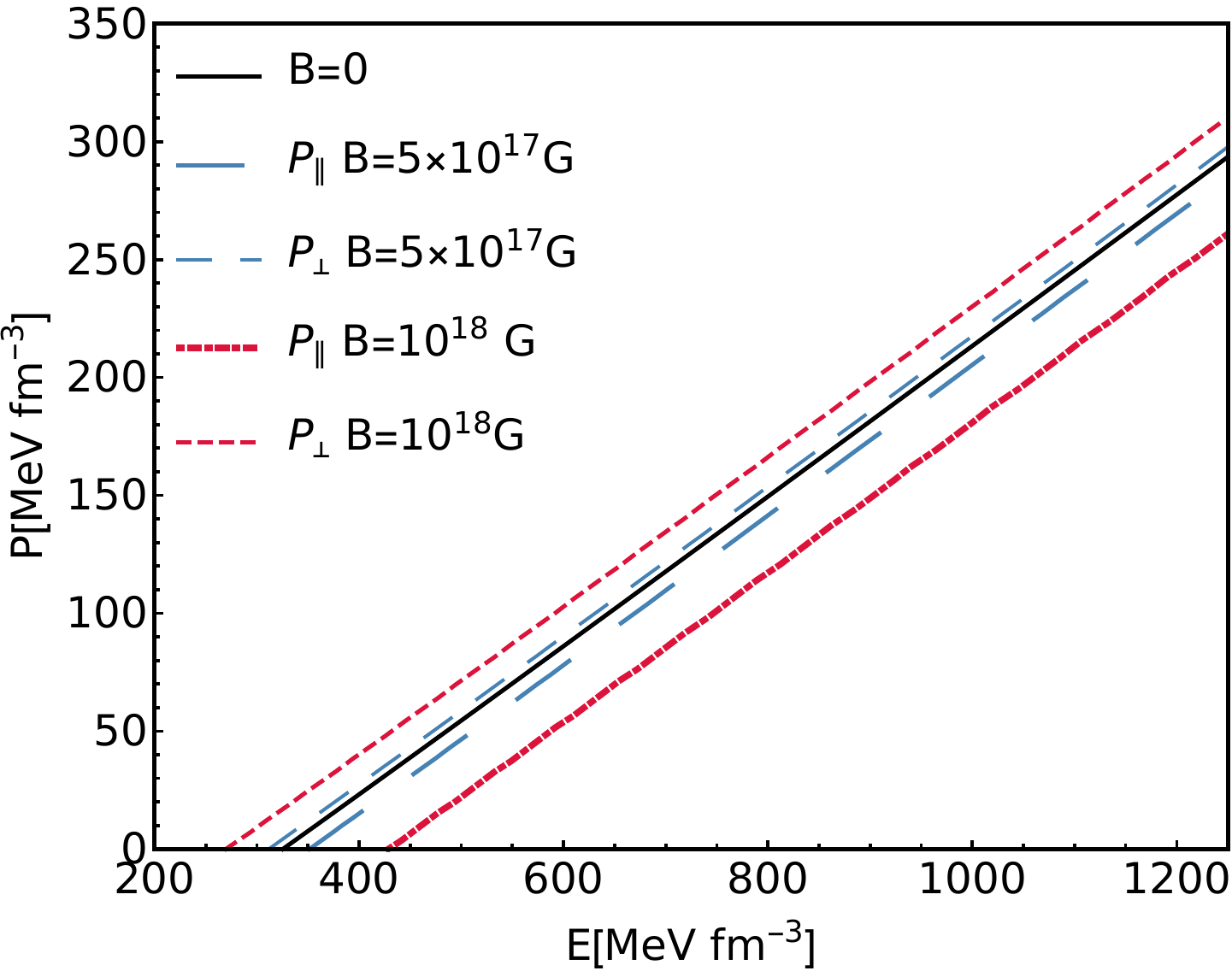}
 	\caption{EoS for magnetized SSs at fixed values of the magnetic field $B=0~$, $B=5\times 10^{17}$~G and $B=10^{18}$~G.}    \label{eosplot}
 \end{figure}

 Fig.~\ref{eosplot} shows Eqs. \eqref{EoSSS} at $B=0$, $B=5\times10^{17}$~G and $B=10^{18}$~G \citep{Daryel2014}. Note that at higher values of the magnetic field, the difference between the perpendicular and parallel pressures is more appreciable. 

 \section{MAGNETIZED Strange stars TOV SOLUTIONS}\label{sec3}
 
 A first step to evaluate the impact of anisotropic pressures in the  SSs structure is to use the magnetized EoS, obtained in the previous section, to solve standard isotropic TOV equations for each pressure component separately. TOV equations are
 \begin{subequations}\label{tov}
 	\begin{eqnarray}
 	&& \frac{dM}{dr}=4\pi Er^2,\\
 	&& \frac{dP}{dr}=-\frac{(E+P)(4\pi Pr^3+M)}{r^2(1-\frac{2M}{r})},
 	\end{eqnarray}
 \end{subequations}
 where $M(r)$ is the mass enclosed in a spherical shell of radius $r$ inside the star. In order to  obtain the sequence of stable stars given a specific EoS, Eqs. \eqref{tov} are integrated until the condition $P(R)=0$ is achieved, being $R$ the radius of the star.
 
\begin{figure}[!h]
 	\includegraphics[width=0.48\textwidth]{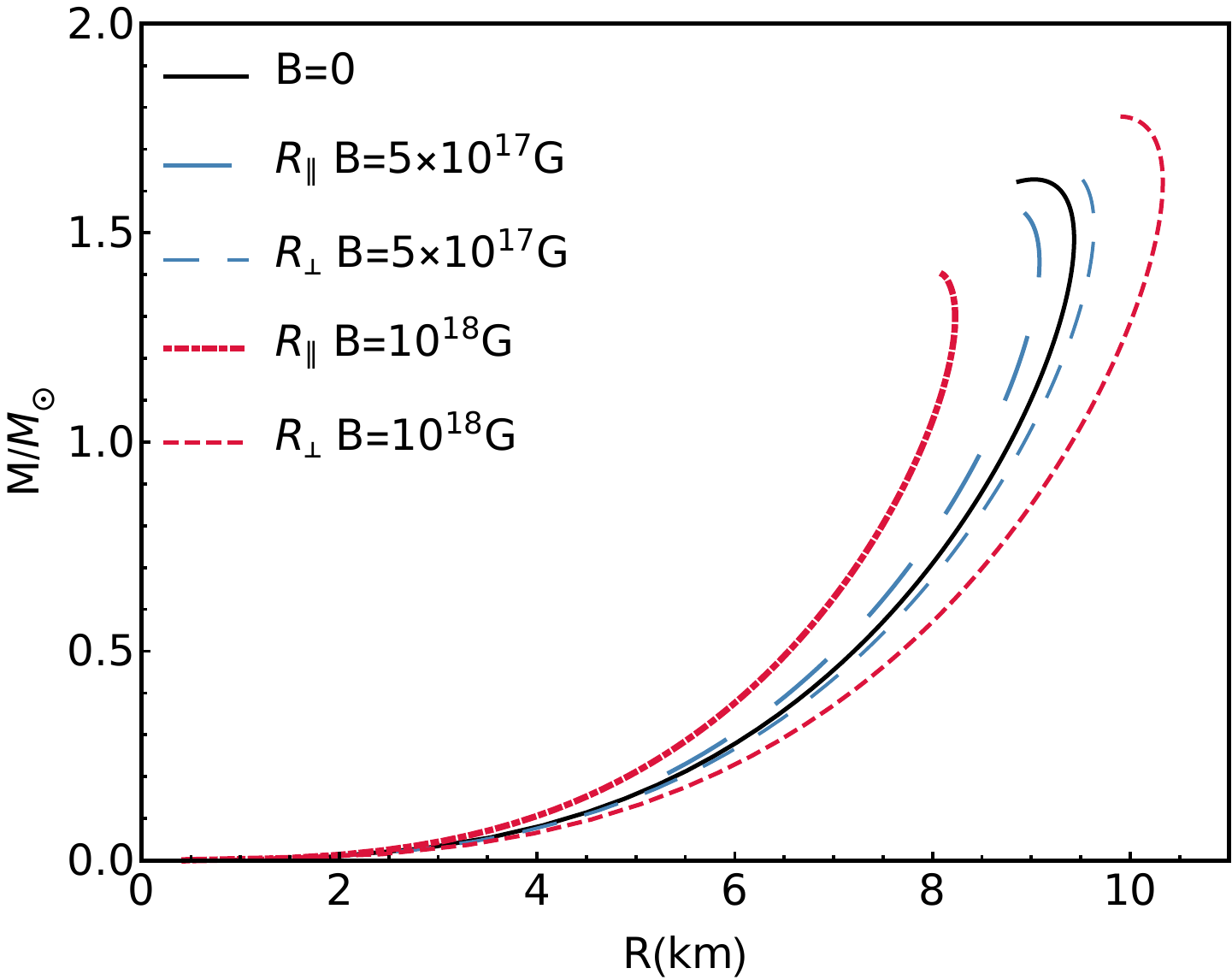}
 	\caption{Isotropic TOV equations solutions for the perpendicular and parallel pressures independently at $B=0~$, $B=5\times 10^{17}$~G and $B=10^{18}$~G.}    \label{tovplot}
 \end{figure}

 In Fig.~\ref{tovplot} we present the mass-radii curves obtained considering the pairs $(E, P_{\parallel})$ and $(E, P_{\perp})$ as independent EoS. The  non-magnetized curve is also plotted for comparison \citep{Daryel12015}. Using one pressure or the other leads to different mass-radii relations, whose differences increase with the magnetic field. 
 
 As we can see in Fig.~\ref{tovplot}, higher pressures give bigger and more massive stars. Also, for a given mass, the difference in the stars size is larger for heavier stars. This suggests that more massive stars will have a greater deformation.
 
 \section{$\gamma$-metric structure equations}\label{sec4}
 
 TOV equations, derived from  spherical symmetry, are compatible with anisotropic EoS with different tangential and radial pressure\footnote{Actually, isotropic EoS is the simplest assumption to obtain hydrostatic equilibrium equation.}\citep{Gleiser}. 
 Nevertheless, the magnetic anisotropy can not accommodate to the spherical symmetry. So, it is an imperative to derive structure equations within axial symmetry. To include such an effect we use the $\gamma$--structure equations obtained in \cite{prd}
 \begin{subequations}\label{gTOV}
 	\begin{eqnarray}
 	\frac{dM}{dr}&=& 4\pi r^{2}\frac{(E_{\parallel}+E_{\perp})}{2}\gamma, \label{gTOV1}\\
 	\frac{dP_{\parallel}}{dz}&=&\frac{1}{\gamma}\frac{dP_{\parallel}}{dr}\nonumber\\
 	&=&-\frac{(E_{\parallel}+P_{\parallel})[\frac{r}{2}+4\pi r^{3}P_{\parallel}-\frac{r}{2}(1-\frac{2M}{r})^{\gamma}]}{\gamma r^{2}(1-\frac{2M}{r})^{\gamma}}, \label{gTOV2}\\
 	\frac{dP_{\perp}}{dr}&=&-\frac{(E_{\perp}+P_{\perp})[\frac{r}{2}+4\pi r^{3}P_{\perp}-\frac{r}{2}(1-\frac{2M}{r})^{\gamma}]}{r^{2}(1-\frac{2M}{r})^{\gamma}}, \label{gTOV3}
 	\end{eqnarray}
 \end{subequations}
 which describe the variation of the mass and the pressures with the spatial coordinates $r,z$ for an anisotropic axially symmetric compact object as long as the parameter $\gamma = z/r =P_{\parallel 0}/P_{\perp 0}$, where $P_{\parallel 0}$ and $P_{\perp 0}$ are the star central pressures,  is close to one.
 
 Note that Eqs. \eqref{gTOV} are coupled through the dependence with the energy density and the mass. When setting B$=0$, the model automatically yields $P_{\perp}=P_{\parallel}$ and $\gamma=1$. This means that we recover the spherical TOV equations from Eqs. \eqref{gTOV} and thus, the standard non-magnetized solution for the structure of COs. Eq. \eqref{gTOV} have been used before to describe magnetized Bose-Einstein Condensate (BEC) stars and white dwarfs \citep{prd,BEC}.
 
\begin{figure}[!h]
 	\includegraphics[width=0.48\textwidth]{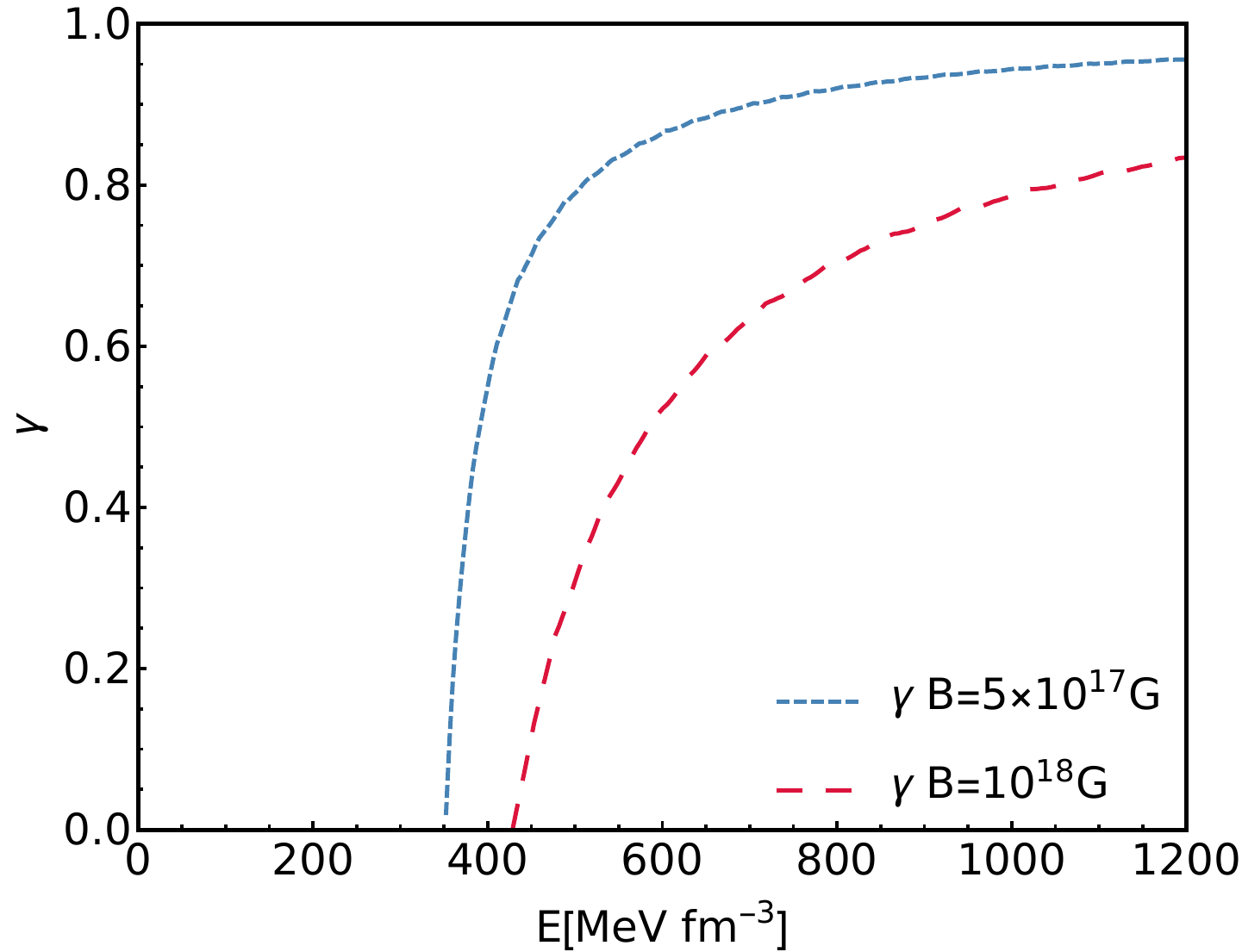}
 	\caption{$\gamma$ parameter as a function of central energy density.}    \label{gammaE}
 \end{figure}

 Since $\gamma \simeq 1$ is a requirement to obtain Eqs. \eqref{gTOV}, we first need to check if they can be used to obtain the mass-radii curve for our EoS. Fig. \ref{gammaE} shows the $\gamma$ parameter as a function of the central energy density. For B$=10^{18}$~G, $\gamma$ is out of the approximation range allowed by the model,  $\gamma\simeq 1$, so this value won't be considered in the next calculations, in which we only use $B=5 \times 10^{17}$~G. 
 
 \subsection{Magnetized Strange Stars numerical results}\label{sec4.1}
 
\begin{figure}[!h]
 	\includegraphics[width=0.48\textwidth]{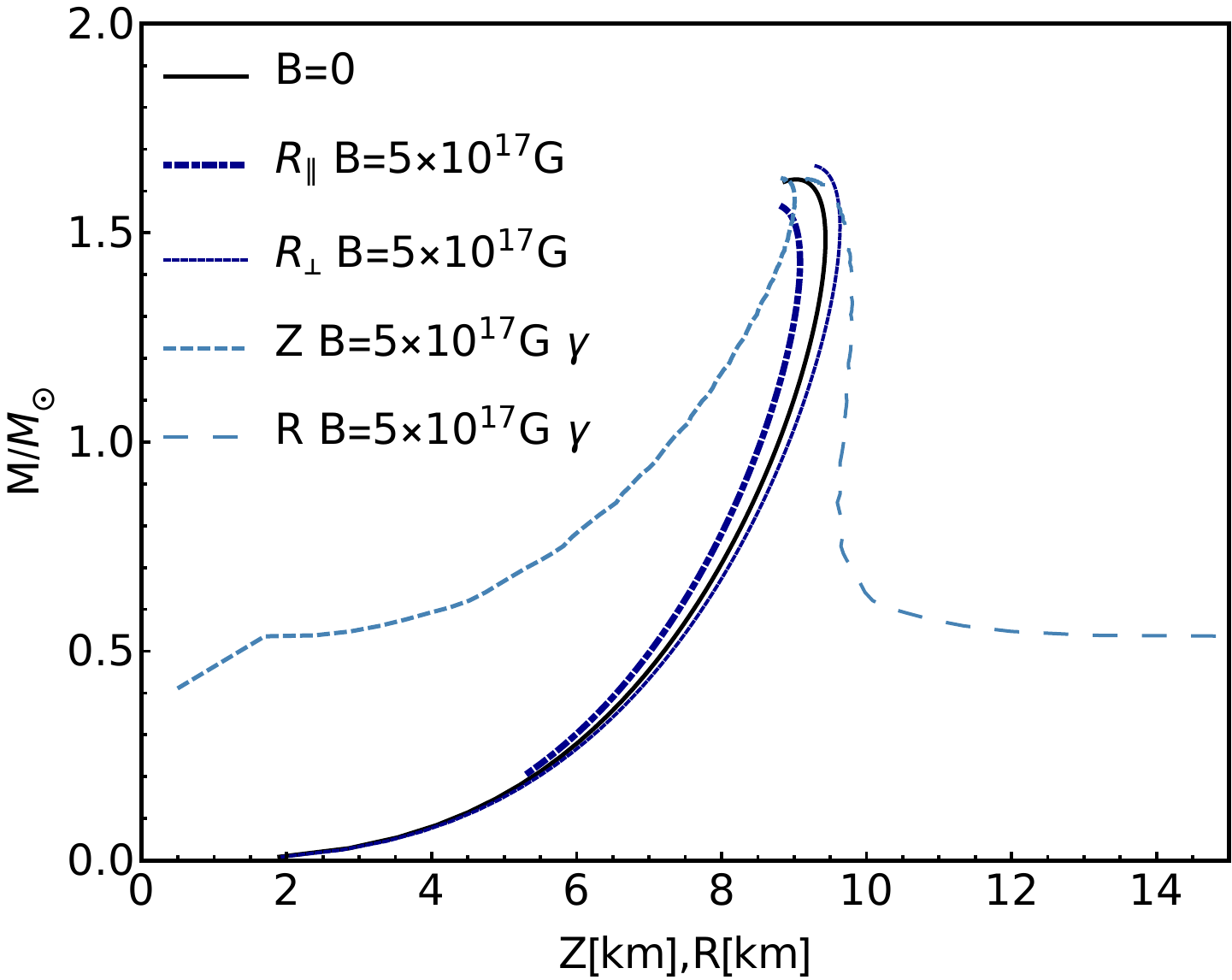}
 	\includegraphics[width=0.48\textwidth]{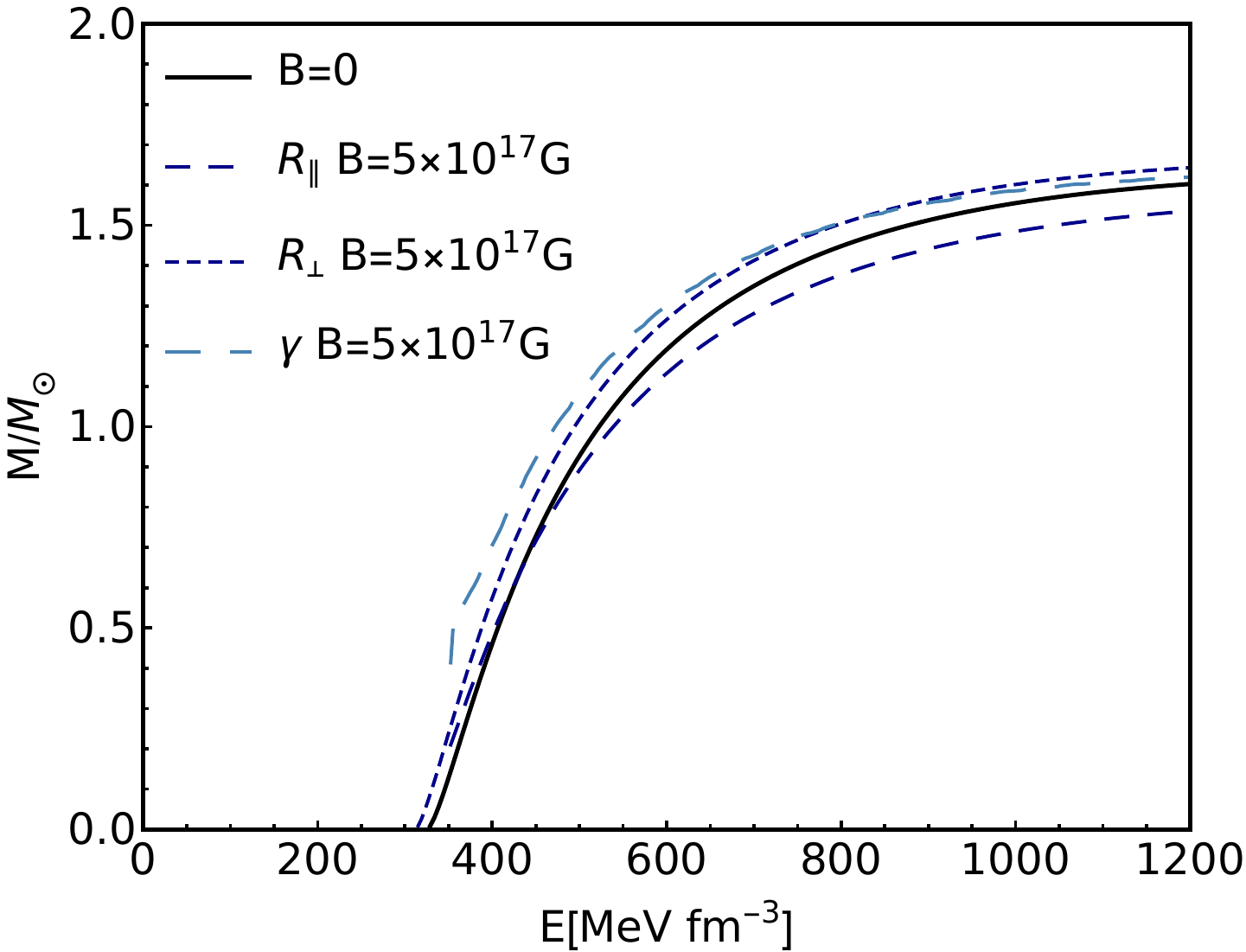}
 	\caption{Solutions of Eqs. \eqref{gTOV}. In the upper panel Mass vs $R$,$Z$. In the lower panel Mass as a function of central energy densities.}
 	\label{gammaMR}
 \end{figure}

 In this section we compare the results of integrating Eqs. \eqref{gTOV} for the EoS of magnetized SSs with those obtained in Sec. \ref{sec3} with TOV equations. Fig. \ref{gammaMR} displays the solutions of Eqs. \eqref{tov} and Eqs.\eqref{gTOV} for $B=5\times10^{17}$~G compared to the non-magnetized solution. Unlike TOV equations, Eqs. \eqref{gTOV} allow us to model the star as a spheroidal with an equatorial radius $R$ and a polar radius $Z$, i.e. the $M-R$ and $M-Z$ curves correspond to an unique sequence of stable stars, while the $M-R_{\perp}$ and the $M-R_{\parallel}$ ones stand for two different sequences.
 
 From the upper panel of Fig. \ref{gammaMR} we see that the stars obtained with Eqs. \eqref{gTOV} are oblate objects ($R>Z$), as expected from TOV solutions, where $R_{\perp}>R_{\parallel}$. However, on the contrary of what happens with TOV solutions, for which the difference between $R_{\perp}$ and $R_{\parallel}$ increases with the mass, the deformation of $\gamma$--structure equations solutions -the distance between $R$ and $Z$- decreases with the mass.
 
 Note also from the lower panel in Fig.~\ref{gammaMR} that solutions of Eqs. \eqref{gTOV} give higher masses than the non-magnetized case for all central densities. This is a particular feature of SSs since for other magnetized compact objects studied with Eqs. \eqref{gTOV} the maximum mass decrease with $B$ \citep{prd,BEC}. The increase of the star`s mass with the magnetic field might be related to the EoS and/or to the $\gamma$--structure equations. The study of this issue is in progress. 
 
 \section{Conclusions}\label{sec6}
 
 We have studied the structure of magnetized strange stars by solving TOV equations for each pressure, $P_{\perp}$ and  $P_{\parallel}$, independently, and a generalized structure set of equations, proposed in \citep{prd}. Since TOV equations can not take into account the joint effect of the anisotropic pressures, their solutions only give a rough idea of how the mass-radii curves of magnetized SSs look like. On the contrary, the $\gamma$--structure equations use both pressures simultaneously, but they only admit small deformations yielding a mathematical lower bound for the central density of the stars that can be solved with these equations.
 
 We would like to remark that each set of structure equations predicts a different qualitative behavior of the deformation induced by the presence of the magnetic field. This is evidence of the strong model-dependence of the observables (mass and radii) and highlights the importance of constructing still more realistic models.
 
 Another important result is that the magnetic field presence augments the stars' maximum masses. When comparing the obtained results with the observations, it can be noticed that the maximum masses
 do not reach the $2\,M_{\odot}$ value. However, the model can be improved by considering a dependence of the bag parameter with the magnetic field and adding rotation to the system.
 
 \section*{Acknowledgments}
 
 The authors have been supported by  PNCB-MES Cuba No. 500.03401 and by the grant of the ICTP Office of External Activities through NT-09. D.A.T. acknowledges the support of the IMPRS-MPSSE hosted by MPI-PKS in Dresden.
 

\begin{thebibliography}{100}
 	
 	\bibitem{Bodmer1971} Bodmer, A. R. 1971, Phys. Rev. D, 4, 1601.
 	
 	\bibitem{ Witten1984} Witten, E. 1984, Phys. Rev. D, 30, 272.
 	
 	\bibitem{Itoh} Itoh, N. 1970, Progress of Theoretical Physics, 44, 291.
 	
 	\bibitem{Drake:2002bj}	Drake, Jeremy J. $\&$ others, Is RXJ1856.5-3754 a quark star?, Astrophys. J, 572, 996-1001, 2002.
 	
 	\bibitem{1992ApJ...392L...9D}  Duncan, R. C. $\&$ Thompson, C.  Formation of Very Strongly Magnetized Neutron Stars: Implications for Gamma-Ray Bursts, apjl, 392, L9, 1992.
 	
 	\bibitem{1998Natur.393..235K} Kouveliotou, C. $\&$ Dieters, S. $\&$ Strohmayer, T. $\&$
 	van Paradijs, J. $\&$ Fishman, G.~J. $\&$ Meegan, C.~A. $\&$
 	Hurley, K. $\&$ Kommers, J. $\&$ Smith, I. $\&$ Frail, D. $\&$
 	Murakami, T., An X-ray pulsar with a superstrong magnetic field in the soft {\ensuremath{\gamma}}-ray repeater SGR1806 - 20, nat, 393, 235-237, 1998,  doi = 10.1038/30410.
 	
 	\bibitem{Daryel22015} Paret, D. M., Horvath, J. E., $\&$ Mart\'inez, A. P. 2015b, Res. Astron.
 	Astrophys., 15, 1735.
 	
 	\bibitem{Lattimerprognosis}
 	J.~M. {Lattimer} and M.~{Prakash}, Phys. Rept. {442} (2007) 109, doi:10.1016/j.physrep.2007.02.003, 	arXiv:astro-ph/0612440.
 	
 	\bibitem{Chaichian}  Chaichian, M., Masood, S. S., Montonen, C., Mart\'inez, A. P., $\&$
 	Rojas, H. P. 2000, Phys. Rev. Lett, 84, 5261.
 	
 	\bibitem{Felipe:2008cm} Felipe, R. G., $\&$ Mart\'inez, A. P. 2009, J. Phys., G36, 075202.
 	
 	\bibitem{prd} Terrero, D. A., Mederos, V. H., P\'erez, S. L., Paret, D. M., Mart\'inez,
 	A. P., $\&$ Angulo, G. Q. 2019, Phys. Rev. D, 99, 023011.
 	
 	\bibitem{Chodos} Chodos,A., Jaffe, R. L., Johnson, K., Thorn, C. B.,$\&$ Weisskopf, V. F.
 	1974, Phys. Rev. D, 9, 3471-3495.
 	
 	\bibitem{Felipe:2007vb} Felipe, R. G., $\&$ Mart\'inez, A. P., $\&$ Rojas, H. P., 2009, Magnetized strange quark matter and magnetized strange
 	quark stars, Phys. Rev., C77, 015807, 2008.
 	
 	\bibitem{Schwinger51} Schwinger, J. 1951, Phys. Rev, 82, 664.
 	
 	\bibitem{Shapiro} S. L. Shapiro y S. A. Teukolsky. {\em Black holes, white dwarfs, y neutron stars: The physics of compact objects} (1983).
 	
 	\bibitem{Camenzind} M. Camenzind. {\em Compact Objects in Astrophysics: White Dwarfs, Neutron Stars y Black Holes.} Astronomy y Astrophysics Library. Springer Berlin Heidelberg (2007). ISBN 9783540499121.
 	
 	\bibitem{Daryel2014} Paret, D. M., Horvath, J. E., $\&$ Mart\'inez, A. P. 2014, Submited to
 	Classical and Quantum Gravity.
 	
 	\bibitem{Daryel12015} Paret, D. M., Horvath, J. E., $\&$ Mart\'inez, A. P. 2015a, Res. Astron.
 	Astrophys, 15, 975.
 	
 	\bibitem{Gleiser}
 	K.~Dev and M.~Gleiser,  Gen. Rel. Grav, 34, 1793, 2002. 
 	doi:10.1023/A:1020707906543, arXiv: astro-ph/0012265.
 	
 	\bibitem{BEC} Angulo, G. Q., Mart\'inez, A. P., Rojas, H. P., $\&$ Paret, D. M. 2019,
 	Int. J. Mod. Phys, D28, 1950135.
 	
 	
 \end{thebibliography}

\end{document}